\documentclass[twocolumn,10pt]{article}

\usepackage[utf8]{inputenc}
\usepackage[T1]{fontenc}
\usepackage{lmodern}
\usepackage{microtype}
\usepackage[margin=0.75in,columnsep=0.28in]{geometry}
\usepackage{amsmath}
\usepackage{amssymb}
\usepackage{booktabs}
\usepackage{array}
\usepackage{graphicx}
\usepackage{xcolor}
\usepackage{tikz}
\usetikzlibrary{positioning,arrows.meta}
\usepackage{float}
\usepackage[numbers,sort&compress]{natbib}
\usepackage[breaklinks,colorlinks,linkcolor=blue!60!black,citecolor=blue!60!black,urlcolor=blue!60!black]{hyperref}

\usepackage[font=small]{caption}

\newcommand{\code}[1]{\texttt{#1}}
\newcommand{\x}{$\times$}
\begin{document}
\twocolumn[
\vspace*{-22pt}
\begin{center}
{\LARGE\bfseries\setlength{\baselineskip}{22pt} Why Git Is the Memory Solution\\ for the Agentic Development Lifecycle\par}
\vspace{8pt}
{\normalsize\itshape Git-bound, routed memory for coding agents: structure, episode, and synthesis under a token budget\par}
\vspace{12pt}
{\normalsize Frank Guo\par}
\vspace{2pt}
{\small Rekal --- \href{https://rekal.dev}{rekal.dev} \,\textperiodcentered\, \href{mailto:guocongmit@gmail.com}{guocongmit@gmail.com} \,\textperiodcentered\, \href{https://github.com/rekal-dev/rekal-cli}{github.com/rekal-dev/rekal-cli}\par}
\end{center}
\vspace{6pt}
\begin{center}
\begin{minipage}{0.92\textwidth}
\small\noindent\textbf{Abstract.}
Coding agents now produce a growing share of a team's code, while the reasoning behind each change --- the alternatives weighed, the constraints discovered, the approaches rejected --- is trapped in assistant transcripts that vanish with the session. Memory for this setting, the agentic development lifecycle (ADLC), is usually posed as one retrieval problem and built as machinery: tiered stores, memory graphs, compiled wikis, model-judged admission. We argue memory should instead be \emph{git-bound} --- built into the repository's version control, inheriting the guarantees the machinery struggles to construct: ground truth from commits, freshness from rebuild, verification from the merge, containment from review. On this ledger we solve two problems separately, then combine them. \emph{Seed supply} is closed as an eight-corpus retrieval study under a pre-registered ship discipline: five imported ranking mechanisms rejected, two kept, and a best configuration of $\approx$0.31 pooled MRR --- $\approx$60\x{} the raw-transcript grep floor, $\approx$15\x{} an honest parsed-turn floor. \emph{Answer assembly} is where ranking stops helping: single-shot retrieval scores only 0.07--0.20 answer-sufficiency on real developer questions, and ungated episode injection measurably degrades good answers. A \emph{router} dispatches breadth to a git-anchored structural map, pointed lookups to confidence-gated episodes, and rationale to \emph{decision synthesis}, which reconstructs why-arcs no single session contains (0.83 sufficiency on a young $\approx$50k-LOC production system). Routed, the system answers at \emph{382--980 tokens per question} --- three orders of magnitude below the recorded history. Because ground truth is mined from commit--session links rather than annotated, every result is replicable on any user's own history at zero labeling cost. The remaining constraint is capture. Code, benchmark, and paper source: \href{https://github.com/rekal-dev/rekal-cli}{github.com/rekal-dev/rekal-cli}.
\end{minipage}
\end{center}
\vspace{1.2em}
]

\section{The question an agent actually asks}\label{sec:question}

Code has a ledger; intent does not. Version control records every line a team ships, but the reasoning that produced those lines --- explored designs, rejected alternatives, the correction a reviewer shouted mid-session --- lives in AI-assistant transcripts that expire with the terminal window. An agent that cannot remember what its own team already tried will confidently re-propose it.

We call this setting the \emph{agentic development lifecycle} (ADLC): git still runs the code's lifecycle, but a growing share of the intent behind each change is produced in agent sessions git never sees. The title's claim is the paper's thesis, argued and then measured: the memory solution for the ADLC is git itself --- \emph{bound}, not bolted on. Bound, because the lifecycle's own primitives supply what memory machinery cannot (ground truth, freshness, verification, containment; \S\ref{sec:guarantees}). And routed, because once the ledger exists, answering from it is not one retrieval problem (\S\ref{sec:modes}).

Start from what memory \emph{is} for an agent. The context window is the scarce resource, so memory is not a store --- it is \emph{context assembly under a token budget}: for each question, deliver the few thousand recorded tokens that change the answer, out of millions \citep{rise2026}. The dominant framing then takes one more step that we refuse: it treats ``each question'' as one retrieval problem, embeds the corpus, ranks sessions, and scores success by Mean Reciprocal Rank against a single gold session.

Watch what developers actually ask, and the framing breaks into three kinds:
\begin{itemize}\itemsep2pt
  \item \textbf{Breadth} (``how is the data-processing pipeline architected end-to-end?''). The answer is spread across many sessions and the code tree; no single episode contains it. Episodic recall floors here by construction.
  \item \textbf{Pointed} (``which session implemented the validation layer, and how?''). The case retrieval is built for --- a specific episode answers.
  \item \textbf{Rationale} (``why was the batch execution engine chosen instead of the managed inference service?''). The answer is a \emph{decision that evolved} across sessions and was never written down in one place. Single-shot retrieval returns one fragment of an arc.
\end{itemize}

These kinds need different mechanisms, and a system that ships only the middle one fails most of the population. The paper therefore solves \emph{two problems, separately, then combines them}. Problem one is \emph{seed supply}: can the right prior sessions be found at all? We close it as a retrieval study --- honest floors, a disciplined mechanism sweep, two validated levers (\S\ref{sec:seed}). Problem two is \emph{answer assembly}: given seeds, can the question actually be answered within a budget? We close it with three modes and a router (\S\ref{sec:modes}--\ref{sec:eval}). The combination --- one engine, one skill-layer router --- outperforms every single-mechanism alternative on coverage of the question kinds, at a fraction of the token cost of the strongest single mode.

\paragraph{Contributions.}
(1) The \emph{git-bound} position: a memory engine built into version control whose guarantees --- annotation, staleness, self-confirmation, contamination --- are inherited by construction (\S\ref{sec:guarantees}), and a self-labeling benchmark that construction makes possible (\S\ref{sec:groundtruth}).
(2) A \emph{closed seed stage}: an eight-corpus retrieval study under a strict incumbent-versus-candidate discipline \citep{rho2026} that establishes honest grep floors, rejects five imported ranking mechanisms with confidence intervals, validates two (per-corpus tuning; a facet term ported from SPM \citep{spm2026}), and derives testable rules for when each helps (\S\ref{sec:seed}).
(3) The \emph{routed three-mode architecture} --- structural map, gated episodes, decision synthesis, shipped as one binary plus one internally-routed agent skill --- and an \emph{answer-sufficiency} evaluation showing each mode wins a different question kind, ungated injection harms, and synthesis reconstructs decision rationale that no retrieval variant reaches (\S\ref{sec:modes}--\ref{sec:eval}).
(4) The token-economics reading: full-kind coverage at 382--980 tokens per routed question (\S\ref{sec:economics}).

\section{A worked example}\label{sec:example}

On Monday, an engineer and an agent rework webhook delivery. Mid-session the engineer interrupts: \emph{``don't retry on a fixed delay --- it stampedes the downstream on recovery.''} The agent switches to exponential backoff with jitter; the change merges. The post-commit hook captures the session --- turns, tool calls, the interruption tagged \code{human\_steering} --- scrubs secrets, and appends it to the repo's ledger with a link to the commit SHA. Nobody writes documentation.

On Friday, three different questions arrive, and the router sends each to a different place. \emph{``Should webhook retries use a fixed delay?''} is pointed: episodic recall returns the Monday session with the steering turn as the top snippet, confidence clears the gate, and a bounded drill reads a five-turn neighborhood --- about 1{,}500 tokens. \emph{``How does delivery work end-to-end?''} is breadth: the structural map --- a condensed subsystem map generated from the repository at its current SHA --- answers from structure; no episode is fetched. \emph{``Why exponential backoff instead of a delivery queue?''} is rationale: a decision-scoped gather collects the decision-relevant turns across sessions (the steering turn, the alternatives discussed before it, the constraint that forced the change) and one synthesis call reconstructs the arc, citing each claim's turn and commit. Same ledger, three mechanisms, each bounded to what the question warrants.

\section{Git-bound: the inherited guarantees}\label{sec:guarantees}

Prior memory systems --- tiered stores \citep{adamem2026}, memory graphs \citep{mragent2026}, compiled wikis \citep{llmwiki2026}, guarded by model-judged admission \citep{edv2026} and evaluated on annotated benchmarks \citep{locomo2024} --- import four problems they then cannot discharge. In the software-engineering setting each has an answer that is a git primitive every team already runs, \emph{by construction}:

\begin{enumerate}\itemsep2pt
  \item \textbf{Annotation} --- where does supervision come from without human labels? $\to$ The \emph{commit}: a post-commit hook links sessions to the verified change they produced; ground truth is mined, not annotated (\S\ref{sec:groundtruth}).
  \item \textbf{Staleness} --- how does derived memory stay current? $\to$ \emph{Rebuild and diff}: the index is disposable and regenerated from the append-only ledger; compiled structure (including \S\ref{sec:modes}'s map) is a function of the tree at a SHA, refreshed by diffing, its drift surfaced as a reviewable \code{git diff}.
  \item \textbf{Self-confirmation} --- who verifies what enters shared memory, if not the model judging its own homework? $\to$ The \emph{merge}: only checkpoints whose commit landed on the default branch are exportable; review + CI + merge are the external verifier.
  \item \textbf{Contamination} --- how does memory cross a scope boundary? $\to$ \emph{Code review} is the sole audited egress; cross-repo memory is index-only and structurally unpushable \citep{memsecurity2026,statecontam2026}.
\end{enumerate}

We take annotation as given by the self-labeling method below and focus the empirical work on the three-mode architecture and its routing. The long-form defense of the four guarantees is in the superseded substrate report (repository history); the design survives here as the platform the modes stand on.

\section{The ledger and the engine}\label{sec:engine}

Rekal is one binary embedding its database engine, embedding models, and compression dictionary; git is the only wire. A post-commit hook parses the active assistant session(s) --- adapters cover four agent CLIs --- deduplicates turns, scrubs secrets and anonymizes paths \emph{before any byte is stored} \citep{statecontam2026}, and appends to an append-only ledger (\code{data.db}, committed via a per-author orphan branch under the merged-only gate). A derived index (\code{index.db}) is rebuilt locally. The ledger records what was \emph{said} --- conversation turns with high-signal roles preserved (\code{human\_steering} corrections, out-of-band compaction \code{summary} distillations) --- and what was \emph{touched}: tool names, file paths, command output, and the commit SHA. Crucially, it does not store diff content: code content is reconstructed from the commit SHA on demand, so intent lives in the ledger and content lives in git (thin on the wire; the division of labor \S\ref{sec:capture} defends).

Recall scores sessions by a weighted hybrid over the parsed turns --- BM25 full-text, latent-semantic, and neural cosine --- with per-role boosts (steering, summary), a subagent down-weight, and a \emph{facet} term (\S\ref{sec:seed}). All weights apply at query time; no reindex to change them. Ship defaults: layer mix 0.35/0.10/0.55, steering boost 1.3, summary boost 1.15, subagent 0.7, max-norm, facet 0.3 (the held-out tuned value; the layer fails soft on corpora without facet material, and an explicit 0 restores the byte-identical pre-facet engine).

\section{Problem one: seed supply, closed}\label{sec:seed}

This section reports the retrieval program to its end and states what it does and does not buy. Protocol throughout: self-labeled gold (\S\ref{sec:groundtruth}), 10\% dev / 90\% held-out test, and the incumbent-versus-candidate discipline of \citet{rho2026} --- a candidate configuration ships only if it beats the incumbent on the held-out split under a paired bootstrap CI excluding zero.

\paragraph{Corpora.} Eight real working corpora spanning four workload classes, anonymized: \emph{Corpus A} (a documentation/knowledge base of $\approx$4{,}000 markdown documents; $\approx$1{,}400 captured sessions; prose-heavy, diverse tooling; retrieval $n_{\text{test}}=212$) and \emph{Corpus B} (a production software system of $\approx$50k lines of code; $n_{\text{test}}=456$--$521$ by split) are the two with statistically clean splits; six smaller corpora (code, docs/architecture, notes, build-scripts, two mixed) are reported directionally and flagged where tuning would overfit.

\subsection{The floors: parsing is most of the miracle}

On Corpus A, term-frequency grep over the \emph{raw} transcript JSONL scores 0.005 pooled MRR. Even at shipped defaults the hybrid over the parsed ledger beats it 37\x{} (0.187; 57\x{} on nDCG@10, non-overlapping bootstrap CIs); under the \emph{best} held-out configuration (\S\ref{sec:mechanisms}, $\approx$0.31 --- derived from the tuned baseline plus the facet marginal) the gap is \emph{$\approx$60\x}. Raw transcripts are adversarial retrieval material (tool dumps, base64, duplicated sidechains); parsing the history into attributed turns is what makes it searchable at all. That result alone, however, is a floor against weak material --- so we also report the stronger \emph{parsed-turn grep} floor (term-frequency rank over the same turns the index sees) on all eight corpora. The hybrid dominates it everywhere: $0.021 \to 0.225$ ($\approx$11\x; \emph{$\approx$15\x} at the best configuration) on Corpus A, $0.028 \to 0.159$ ($\approx$6\x) on Corpus B, and on every small corpus besides (floors 0.08--0.19 vs hybrids 0.21--0.58). Ranking still earns its place on parsed turns; grep does not close the gap once parsing is granted.

\subsection{The mechanism study, and the best configuration}\label{sec:mechanisms}

\begin{table}[t]
\centering\small
\setlength{\tabcolsep}{4pt}
\begin{tabular}{@{}p{0.58\linewidth}cc@{}}
\toprule
\textbf{Seed stage} & \textbf{A} & \textbf{B}\\
\midrule
grep, raw transcript JSONL (floor) & 0.005 & ---\\
grep, parsed turns (honest floor) & 0.021 & 0.028\\
BM25-only & 0.200 & 0.148\\
hybrid, shipped default & 0.225 & 0.159\\
\textbf{best config: tuned hybrid + facet} & \textbf{$\approx$0.31} & \textbf{$\approx$0.24}\\
\midrule
\multicolumn{3}{@{}l@{}}{\emph{Mechanism sweep (paired bootstrap CI; detail in App.~\ref{app:seed})}}\\
\midrule
rejected: RRF / temporal / dilution / z-norm / embedder swap & \multicolumn{2}{r@{}}{n.s.\ or degr.}\\
kept: per-corpus weight tuning & \multicolumn{2}{r@{}}{B +0.032*}\\
kept: SPM facet term \citep{spm2026}, fb 0.3 & \multicolumn{2}{r@{}}{A +0.110*, B +0.053*}\\
\bottomrule
\end{tabular}
\caption{Problem one closed (pooled MRR, held-out test): floors, the mechanism study under the RHO discipline \citep{rho2026}, and the best held-out configuration --- $\approx$15\x{} the honest grep floor on Corpus A. Five imported mechanisms rejected with paired bootstrap intervals (tuning: B +0.032 [.016,.049]; facet marginals: A +0.110 [.056,.173], B +0.053 [.012,.107]); two kept. Best-config cells are derived from the tuned baseline plus the facet marginal (exact cells in the run record). Negative results are load-bearing: they close the stage.}
\label{tab:seed}
\end{table}

Three findings organize Table~\ref{tab:seed}. \textbf{Hybrid beats BM25 where paraphrase opens a surface-form gap and not elsewhere}: pooled on Corpus A the two sit within each other's CIs (0.187 vs 0.171 on the primary label run), the separation is a provenance effect (T1 MRR 0.301 vs 0.248; R@5 0.537 vs 0.425), and the only corpus where the hybrid wins outright is the noisy-commit prose corpus (+0.352 [.158,.549]) --- on exact-vocabulary code corpora it ties.

\textbf{The facet term is the one imported mechanism that survives end-to-end.} SPM \citep{spm2026} derives structured facets (task, data schema, tool config, output constraints) for session-start context assembly; we port the idea to a retrieval term: a deterministic per-session facet document (distinct tool paths + command prefixes + steering text; no LLM, built at index time), BM25-searched as a fourth, orthogonal hybrid term --- additive and config-gated, so at \code{facet\_boost}=0 the engine is byte-identical to the baseline. Root cause of the win: the answer to ``what tools/config did session X use'' lives in tool-call \emph{metadata} that a session's conversational turns often never mention. Bolted naively onto the untuned default mix the term looks corpus-conditional (significant on A, null on B); the joint re-tune --- one never ships an untuned knob --- shows the marginal is significant on \emph{both} corpora, with magnitude monotone in tool-path diversity (Table~\ref{tab:facet}, App.~\ref{app:seed}).

\textbf{The embedded model is not the bottleneck.} Substituting a strong hosted general-purpose embedder, under a protocol that gives it every advantage, moves the neural-only ablation slightly and the shipped hybrid not at all (n.s.\ on both corpora; Table~\ref{tab:embed}, App.~\ref{app:seed}): \emph{alignment, not embedding quality, is the binding constraint} \citep{anms2026}, and the local-first default --- embedding model inside the binary, no API --- costs approximately nothing in recall.

\subsection{What the mechanism study teaches}

Every pure re-ranking mechanism failed; the single retrieval gain came from adding an \emph{orthogonal evidence layer} for a \emph{kind of question} the turn index structurally misses. Gains come from coverage of question kinds, not rank polish. The seed stage is hereby closed --- retrieval is a solved-enough seed supplier with two validated levers and characterized limits --- and the rest of the paper applies the same lesson above retrieval, where the layers are no longer index columns but memory \emph{modes}.

\section{Ground truth without annotation}\label{sec:groundtruth}

No benchmark exists for repo-grounded intent recall: conversational suites \citep{locomo2024} evaluate chat personas; IR suites evaluate document QA. RekalBench's defining property follows from the git binding: \emph{the corpus labels itself.} Every checkpoint records which sessions produced which commit; a SQL miner over ledger plus git topology emits gold pairs with no human in the loop --- provenance (commit $\to$ producing session), decision recall (steering turns), dead ends (never-merged branches), multi-hop (file-co-occurrence session pairs). An LLM paraphrases each label's context into a natural question under a 4-gram Jaccard $\le 0.30$ leakage ceiling. Supervision cost: zero. The retrieval study of \S\ref{sec:seed} runs on these labels (415 pairs on Corpus A alone; T4 multi-hop supplies 92--104 pairs each in the three high-co-occurrence corpora); label noise biases scores \emph{down}, not up, so the numbers are conservative. Because labels are mined, the entire harness is public, fully local, and runnable by anyone on their own store.

\section{Problem two: answer assembly --- three modes and a router}\label{sec:modes}

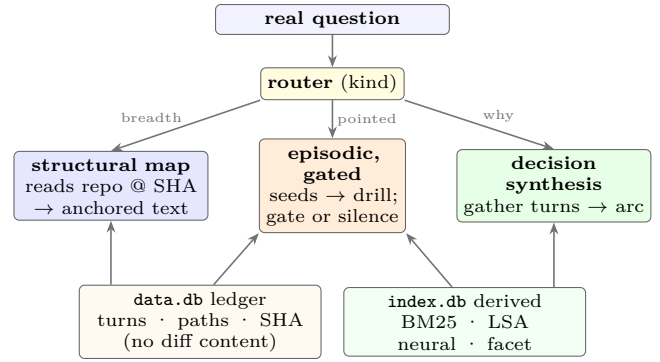
\begin{figure}[t]
\centering
\begin{tikzpicture}[
  font=\scriptsize,
  box/.style={draw=black!60, rounded corners=2pt, align=center, inner sep=2.5pt},
  arr/.style={-{Stealth[length=4.5pt]}, black!55, semithick}
]
\node[box, fill=blue!5, minimum width=88pt]  (q)   at (0, 0)      {\textbf{real question}};
\node[box, fill=yellow!15]                   (r)   at (0, -0.85)  {\textbf{router} (kind)};
\node[box, fill=blue!10,   text width=68pt]  (map) at (-2.93, -2.2) {\textbf{structural map}\\ reads repo @ SHA\\ $\to$ \mbox{anchored text}};
\node[box, fill=orange!15, text width=50pt]  (ep)  at (0, -2.2)     {\textbf{episodic, gated}\\ seeds $\to$ drill;\\ gate or silence};
\node[box, fill=green!10,  text width=68pt]  (syn) at (2.93, -2.2)  {\textbf{decision synthesis}\\ gather turns $\to$ arc};
\node[box, fill=orange!5,  text width=86pt]  (data) at (-1.75, -4.0) {\code{data.db} ledger\\ turns \textperiodcentered{} paths \textperiodcentered{} SHA\\ (no diff content)};
\node[box, fill=green!5,   text width=86pt]  (idx)  at (1.75, -4.0)  {\code{index.db} derived\\ BM25 \textperiodcentered{} LSA\\ neural \textperiodcentered{} facet};
\draw[arr] (q) -- (r);
\draw[arr] (r.south west) -- node[above left, inner sep=1pt]{\tiny breadth} (map.north);
\draw[arr] (r.south) -- node[right, inner sep=1.5pt]{\tiny pointed} (ep.north);
\draw[arr] (r.south east) -- node[above right, inner sep=1pt]{\tiny why} (syn.north);
\draw[arr] (data.north -| map.south) -- (map.south);
\draw[arr] (data.70) -- (ep.south west);
\draw[arr] (idx.110) -- (ep.south east);
\draw[arr] (idx.north -| syn.south) -- (syn.south);
\end{tikzpicture}
\caption{One git-native ledger supplies seeds and structure; a router sends each question to one of three modes by kind. Code content is reconstructed from the commit SHA on demand, so the ledger stays thin. Episodes are confidence-gated before they join the map.}
\label{fig:arch}
\end{figure}

\subsection{Structural map (breadth)}

The map is a condensed subsystem description \emph{authored by an LLM that reads the repository} --- its directory skeleton, its own README/architecture documents --- not clustered from co-occurrence statistics (we tried; the statistical version grouped tool-call-id dumps and scratch files into meaningless clusters; comprehension filters what statistics cannot). Its reader is an agent, so its format is structured text with greppable path anchors rather than a diagram: subsystems cut by responsibility with purpose stated as behavior, load-bearing edges labeled with what crosses them, and the main flows --- dense per token, and section-scoped so refreshes rewrite only what changed. The map is watermarked with the HEAD commit and regenerated on demand, so it is never a stale snapshot: it is a function of the tree at a SHA, refreshable by diffing only the sections whose files changed --- the staleness guarantee of \S\ref{sec:guarantees} applied to compiled structure. It answers ``what exists and how it connects,'' and nothing about ``why.''

\subsection{Episodic recall, confidence-gated (pointed)}

For pointed questions the engine returns top-$k$ seed sessions and the agent drills into the cheapest dense anchor (a summary turn, else a turn window). The critical design point is that episodes must be \emph{gated}: we calibrated a hit signal from the retriever's own per-layer scores --- top-1 score and top-1--top-2 gap separate hits from misses (means 0.91 vs 0.87; gap 0.046 vs 0.017) --- and inject episodes only when the signal clears the bar. Ungated, low-confidence episodes \emph{poison} a good map (\S\ref{sec:ablations}).

\subsection{Decision synthesis (rationale)}

For ``why'' questions the agent does not retrieve one session; it \emph{gathers every decision-relevant turn} --- a direct query over the ledger for the choice, its alternatives, and reasoning markers (``because'', ``instead of'', ``constraint'', ``rejected'') --- then \emph{synthesizes the arc}: original design $\to$ alternatives rejected $\to$ the constraint that forced the change $\to$ final rationale, with every claim carrying its turn and commit pointer (the self-confirmation guarantee applied to generated text). Where the reasoning references code, the diff is pulled from the commit SHA at synthesis time. This is the mode single-shot retrieval cannot emulate, because the answer is \emph{distributed} and was never a single record.

\subsection{The router is a skill: gated triage over workflows}\label{sec:router}

The router and its three modes ship as \emph{one} agent skill --- written workflow playbooks, internally routed, driving engine primitives --- so the agent loads a single policy. Its first decision is not which memory mode but \emph{which substrate}: present-tense code facts belong to the tree (grep and read at HEAD) and are deliberately not answered from memory --- a recalled episode about how X \emph{used} to work can mislead about how it works now; past-tense intent belongs to the ledger; structure belongs to the map, which bridges the two (derived from the tree, serving memory). Grep the tree for present tense, recall the ledger for past tense --- and stay silent when it is neither. Within the ledger, a \emph{triage} step then classifies the question's kind from its shape (``how does X work end-to-end'' $\to$ breadth; ``which session did X'' $\to$ pointed; ``why X instead of Y'' $\to$ rationale), a \emph{gate} decides whether episodic evidence enters at all (the confidence signal of \S\ref{sec:modes}.2 --- below the bar, the mode stays silent rather than injecting noise), and each mode is then a \emph{workflow}: map $\to$ regenerate-on-diff and answer from structure; hunt $\to$ seeds, gate, drill the cheapest dense anchor; why $\to$ gather decision turns, pull diffs by SHA, synthesize with pointers. Nothing in this pipeline is a trained component: triage rules, gate thresholds, and workflows are inspectable text, versioned in git like everything else, so improving the routing policy is editing a file under review --- the same maintenance story as the rest of the system. Rationale lives in turns; structure lives in the tree; the modes are \emph{routed, not stacked}.

\section{Evaluation: answer-sufficiency, not rank}\label{sec:eval}

We reject single-gold MRR as the headline. Our metric is \emph{answer-sufficiency}: for a real question, assemble context under a mode, have a distinct blind judge rate whether the context is SUFFICIENT (1), PARTIAL ($\tfrac12$), or INSUFFICIENT (0) to answer, and report the mean with bootstrap 95\% CIs, plus average tokens per question. Ground truth for the corpora is self-labeled; questions are real developer questions tagged \emph{broad / pointed / why} ($n{=}15$ per corpus; per-question judgments and generated maps are in the run directory). Corpus A is the documentation/knowledge corpus of \S\ref{sec:seed} ($\approx$4{,}000 markdown documents); Corpus B is a production software system of $\approx$50k lines of code --- a layered data/ML pipeline (ingest $\to$ transform $\to$ model build) inside a $\approx$3{,}000-session repository --- with uniform tooling and a \emph{young} recorded history: the harder test, memory that has barely accumulated.

\subsection{Each mode wins a different kind}

Table~\ref{tab:suff} is the study's main table. We stress up front that with 15 questions per corpus, a single judge, and wide bootstrap intervals, the overall point estimates are \emph{not} statistically separable --- most CIs overlap. We therefore read the results as directional patterns across question kinds, not as a ranking of modes; the routing signal is in the per-kind columns, not the overall.

\begin{table*}[t]
\centering\small
\begin{tabular}{@{}lccccc@{}}
\toprule
\textbf{Mode} & \textbf{Overall} (95\% CI) & \textbf{Broad} & \textbf{Pointed} & \textbf{Why} & \textbf{Avg tok}\\
\midrule
\multicolumn{6}{@{}l}{\emph{Corpus A --- documentation/knowledge ($\approx$4{,}000 markdown docs, $\approx$1{,}400 sessions)}}\\
structural map & 0.33 (.13--.53) & 0.50 & 0.17 & 0.33 & 201\\
episodic single-shot & 0.20 (.00--.40) & 0.33 & 0.17 & 0.00 & 914\\
routed (map + gated episodes) & 0.43 (.20--.67) & 0.83 & 0.50 & 0.17 & 382\\
decision synthesis & 0.47 (.20--.67) & 0.50 & 0.50 & 0.33 & 2762\\
\midrule
\multicolumn{6}{@{}l}{\emph{Corpus B --- production software system ($\approx$50k LOC, young history)}}\\
structural map & 0.53 (.30--.77) & 0.50 & 0.33 & 1.00 & 969\\
episodic single-shot & 0.07 (.00--.27) & 0.00 & 0.00 & 0.33 & 648\\
routed (map + gated episodes) & 0.60 (.37--.80) & 0.67 & 0.42 & 0.83 & 980\\
decision synthesis & \textbf{0.83 (.67--.97)} & 0.92 & 0.92 & 0.50 & 3135\\
\bottomrule
\end{tabular}
\caption{Answer-sufficiency by mode on both corpora ($n{=}15$ real questions each, tagged broad/pointed/why; blind judge; bootstrap 95\% CI on the overall). With this sample size the overall point estimates carry wide intervals and most pooled contrasts are \emph{not} statistically separable --- read the results as directional patterns across question kinds, not a ranking of modes. The routing signal is in the per-kind columns: episodic-alone floors on breadth and near-floors overall on the uniform young corpus (0.07); the map answers breadth but not why; synthesis dominates the corpus whose decisions are recent. Routing is no worse than the best single mode on every kind and strictly better on breadth, at 382--980 tokens.}
\label{tab:suff}
\end{table*}

\subsection{Gating and the rationale ablation}\label{sec:ablations}

\begin{table}[t]
\centering\small
\begin{tabular}{@{}p{0.44\linewidth}cp{0.30\linewidth}@{}}
\toprule
\textbf{Arm} & \textbf{Result} & \textbf{Note}\\
\midrule
\multicolumn{3}{@{}l}{\emph{Episode gating, isolated (Corpus B, $n{=}12$)}}\\
episodic single-shot & 0.00 & near-floor; uniform pipeline\\
structural map & 0.63 & reads the real layered architecture\\
map + episodes, \emph{ungated} & 0.29 & low-confidence episodes \emph{poison} the map\\
map + episodes, \emph{gated} & 0.50 & gate suppresses 11/12 bad injections\\
\midrule
\multicolumn{3}{@{}l}{\emph{The execution-engine rationale question (sufficient?)}}\\
structural map & partial & structure, not rationale\\
the source code itself & no & uses the engine; no why\\
episodic single-shot (top-3) & no & returns one fragment\\
synthesis, under-gathered (4 terms) & no & too few turns gathered\\
synthesis, adequate gather (30 turns) & \textbf{yes} & blind judge; 2.1k tok\\
\bottomrule
\end{tabular}
\caption{Corpus B ablations. Top: gating isolated on a separate question set --- ungated episodes degrade a good map and the confidence gate recovers most of the loss (0.29 $\to$ 0.50 of the 0.63 map baseline; only 1 of 12 retrievals cleared the gate, so gating here mostly means silence). Bottom: the mode ablation on one evolved architectural decision --- only a decision-scoped gather plus synthesis answers it.}
\label{tab:ablate}
\end{table}

Three lessons. \textbf{Episodes must be gated, or they degrade the map}: naive context-stuffing --- the integration everyone builds first --- is measurably harmful, and knowing when to stay silent is part of the system. \textbf{Decision synthesis is the most distinctive mode}, and its quality is bounded by the \emph{gather}, not the synthesizer: a weak four-term gather starved the same model into INSUFFICIENT; an adequate gather (30 decision-relevant turns, $\approx$2.1k tokens, one synthesis call) reconstructed the full arc for a design that began under one constraint set and drifted to ``good enough.'' The lesson is precise: \emph{the rationale was in memory all along; single-shot retrieval fragments it and a poor gather starves it, but a decision-scoped gather plus synthesis assembles it.} And \textbf{the young corpus is the strong case for memory}: a barely-accumulated history already answers 0.83 of real questions under synthesis --- memory pays off long before it is big.

\subsection{The economics: coverage at cost}\label{sec:economics}

Table~\ref{tab:waterfall} is the paper in one table: each stage added covers a question kind the previous stages missed, and the combined routed system answers at a cost three orders of magnitude under the recorded history --- with cost following the \emph{question}, not the corpus.

\begin{table*}[t]
\centering\footnotesize
\begin{tabular}{@{}p{0.27\textwidth}p{0.26\textwidth}p{0.18\textwidth}p{0.17\textwidth}@{}}
\toprule
\textbf{Stage added} & \textbf{What it adds} & \textbf{Sufficiency (A / B)} & \textbf{Tokens/question}\\
\midrule
grep, raw transcript JSONL & nothing --- the floor (0.005 MRR) & $\approx$0 & unbounded scan\\
grep, parsed turns & weak seeds (0.021/0.028 MRR) & --- & unbounded scan\\
best seed config: tuned hybrid + facet ($\approx$0.31 MRR) & pointed answers; seeds for every mode & 0.20 / 0.07 (single-shot) & $\approx$0.9k; $\approx$1.5k with drill\\
+ structural map & breadth & 0.33 / 0.53 & 201 / 969, amortized regen\\
+ decision synthesis & rationale, multi-hop & 0.47 / \textbf{0.83} & $\approx$2.8--3.1k\\
\textbf{routed: map + gated episodes} & \textbf{all kinds at the per-kind floor} & \textbf{0.43 / 0.60} & \textbf{382 / 980}\\
\bottomrule
\end{tabular}
\caption{Coverage at cost --- the two problems combined. Seed supply makes the ledger findable (grep floors $\to$ $\approx$0.31 best config: $\approx$60\x{} raw, $\approx$15\x{} the honest floor); answer assembly converts findable into answered, kind by kind; routing buys the per-kind floor at 382--980 tokens against a recorded history of tens of thousands of turns. Synthesis buys the hardest kind at $\approx$3\x{} the routed price --- cost follows the question. With the question-kind distribution of real usage (mined from the ledger's own query log; instrumented, future run) the last row becomes one expected-cost figure: $\mathbb{E}[\text{tokens}] = \sum_{\text{kind}} p(\text{kind}) \cdot \text{cost}(\text{mode})$.}
\label{tab:waterfall}
\end{table*}

\section{The real bottleneck is capture, not ranking}\label{sec:capture}

Every failure in \S\ref{sec:eval} that was not a routing error was a \emph{capture gap}: the answer had never been verbalized in the ledger. This reframes the research target. The ledger deliberately keeps only what was said plus what was touched (git SHA; no diff content) --- thin to stay on the wire --- and synthesis pulls code from the commit on demand. Our finding is that the division of labor is correct but incomplete on one side: intent lives in the ledger and content lives in git, so the remaining loss is reasoning that agents never say out loud. Keeping capture thin is therefore not a limitation to walk back --- it preserves the git-bound answers to staleness and contamination that a fatter, content-bearing ledger would compromise --- but capture completeness (verbalization rate, measured per corpus) is the binding constraint on everything above it, and the next unit of improvement per token spent lies there, not in ranking.

\section{Related work}\label{sec:related}

Memory-graph and tiered-store systems \citep{mragent2026,adamem2026,automem2026} and self-evolving retrieval over compiled wikis \citep{llmwiki2026} target long-horizon recall but assume annotation, maintain compiled structure in software, and admit via model judgment --- the four \S\ref{sec:guarantees} problems. The data-management evaluation \citep{anms2026} reaches the verdict this paper takes literally: effectiveness depends on aligning memory structure with the workload; the coding workload's structure is the repository, and Rekal aligns by construction. On the compression axis we side with preservation-with-attribution \citep{emem2025} (raw turns, harvested summaries) over write-time consolidation \citep{adamem2026}; on the retrieval axis with active reconstruction \citep{mragent2026} (the skill drives search--facet--zoom--drill loops); the storage is flat indexes joined at query time \citep{sag2026}. Against retrieval-free direct corpus interaction \citep{dci2026,grepseek2026}, our floors sharpen RISE's argument \citep{rise2026} at the retrieval layer. SPM \citep{spm2026} contributes the structured-facet thesis our seed stage validates in a different role (retrieval term, not context assembly). The retrospective-optimization discipline \citep{rho2026} governs every shipped mechanism. Our departure is threefold: (i) git as the source of ground truth, freshness, and the sharing gate rather than rebuilding them; (ii) single-gold MRR replaced, as the headline, by routed answer-sufficiency; (iii) decision synthesis over the episodic trail identified as a distinct memory mode, not a retrieval variant.

\section{Limitations}\label{sec:limits}

The sufficiency evidence is small: $n{=}12$--$15$ questions per corpus, one blind judge, one execution model; we report CIs and treat magnitudes as directional, and the honest summary of Table~\ref{tab:suff} is per-kind patterns, not separable pooled rankings. Answer-sufficiency is a proxy for task help, not task completion. The map is only as good as the repository's own structure and documentation; synthesis reconstructs rationale only to the extent it was verbalized (\S\ref{sec:capture}); the judge saw no trail --- model and question sets are modest and the harness supports scaling both. The seed-stage study is one operator's corpora; because labels are mined, replication is free for any user on their own history. Specified but not run here: synthesis on the mined T4 multi-hop gold (92--104 pairs per high-co-occurrence corpus), a second judge model with agreement, the wild-question kind-distribution (and with it the expected-cost figure), and gate recalibration after the facet term enters the score mix.

\section{Conclusion}\label{sec:conclusion}

Agent memory is not one retrieval problem but three modes --- structure, episode, and synthesis --- over a ledger whose hard guarantees come from version control. The paper's shape is deliberate: two problems solved \emph{separately} --- seed supply, closed as a retrieval study with honest floors, a mechanism graveyard, two validated levers, and rules for when each helps; answer assembly, closed with three modes behind a skill-router --- and then \emph{combined}, where the combination outperforms every single-mechanism alternative on coverage of the question kinds at a fraction of the strongest mode's token cost (Table~\ref{tab:waterfall}). The value above ranking is in routing each question to the mode that can answer it, gating episodes on confidence so they help rather than harm, and reconstructing evolved decisions by synthesis rather than retrieval. The binding constraint is capture. One tool, three modes, git-bound --- and the strongest result is the one the field has not been measuring: memory can reconstruct \emph{why} a system became what it is. Git already runs the ADLC's code; bound and routed, it runs its memory too.

\paragraph{Reproducibility.}
Every value traces to a committed run record (per-mode sufficiency judgments, blind-judged synthesis runs, generated maps, gating ablation; retrieval matrix, mechanism sweeps, facet screens) --- anonymized aggregates under \code{docs/research/runs/}. Corpora are anonymized (A: a documentation repository; B: a production data/ML pipeline subsystem); no session content or corpus identity leaves the operator's machine --- published artifacts are aggregates, prompts, and code. The router and its three mode workflows ship as one agent skill (\code{rekal}); the judge is a single automated model and the questions are the authors' own corpora --- the study is a within-system characterization, not a competition. Engine, skills, benchmark spec, extraction SQL, and this paper's source: \href{https://github.com/rekal-dev/rekal-cli}{github.com/rekal-dev/rekal-cli} (\code{docs/research/}).

\appendix

\section{Seed-stage detail}\label{app:seed}

\paragraph{The facet term at two operating points.}
Evaluated by bolting it onto a fixed configuration, a mechanism can appear corpus-conditional when it is merely mis-tuned; only the joint re-tune separates ``does not work here'' from ``was not given its operating point.'' What remains genuinely corpus-conditional is the \emph{magnitude}: the marginal is monotone in tool-path diversity (the high-diversity corpus gains +0.110; the uniform-pipeline corpus +0.053, at $\approx$2.3\x{} less path diversity per call) --- a testable deployment rule. The term ships enabled at the tuned \code{facet\_boost} of 0.3; corpora without facet material pay nothing (the facet index is guarded), and an explicit 0 restores the byte-identical pre-facet engine. We test the retrieval port only, explicitly not a reproduction of SPM's task-completion result, which would need a curated gold set.

\begin{table}[H]
\centering\footnotesize
\setlength{\tabcolsep}{4pt}
\begin{tabular}{@{}lcc@{}}
\toprule
\textbf{Facet term} & \textbf{Corpus A} & \textbf{Corpus B}\\
\midrule
screen: facet vs turns & +0.184 [.05,.32]* & +0.016 n.s.\\
end-to-end (fb 0.6) & +0.116 [.04,.20]* & $-$0.019 n.s.\\
joint re-tune (fb 0.3) & +0.110 [.06,.17]* & +0.053 [.01,.11]*\\
\bottomrule
\end{tabular}
\caption{The facet term at two operating points (held-out, paired bootstrap CIs; * = significant). The screen runs facet-document BM25 against turn recall on structural queries; the end-to-end row bolts the term onto the untuned default mix (lifting Corpus A 0.179$\to$0.294); the joint re-tune reports the facet marginal with the mix tuned around it. Naively bolted on, the term looks corpus-conditional; jointly re-tuned --- one never ships an untuned knob --- the marginal is significant on both corpora: the naive null was an artifact of an untuned operating point, not an absent effect. On Corpus A the marginal takes the best held-out hybrid to $\approx$0.31 pooled MRR.}
\label{tab:facet}
\end{table}

\paragraph{Embedding options.}
The substitution protocol gives the alternative its best shot: same held-out split and candidate pool; asymmetric query/document input types sent natively; the content-hash embedding cache rebuilt per model; four configurations per embedder (neural-only and hybrid, each at default and dev-tuned weights); paired per-query bootstrap.

\begin{table}[H]
\centering\footnotesize
\setlength{\tabcolsep}{4pt}
\begin{tabular}{@{}lcc@{}}
\toprule
\textbf{Configuration (pooled MRR)} & \textbf{Corpus A} & \textbf{Corpus B}\\
\midrule
BM25-only & 0.200 & 0.148\\
neural-only --- embedded (local) & 0.080 & 0.045\\
neural-only --- hosted & 0.093 & 0.055\\
hybrid --- embedded, default mix & 0.225 & 0.159\\
hybrid --- embedded, dev-tuned mix & 0.211 & \textbf{0.182}\\
hybrid --- hosted, default mix & 0.217 & 0.140\\
hybrid --- hosted, dev-tuned mix & \textbf{0.229} & 0.170\\
\bottomrule
\end{tabular}
\caption{Embedding options on the held-out split. The hosted embedder (a large general-purpose API model) lifts the \emph{neural-only} ablation slightly on both corpora, but the \emph{hybrid} --- the shipped system --- is flat on A and worse on B (hosted-default 0.140 falls below BM25-only); all hybrid deltas n.s.\ under paired per-query bootstrap. Scope: one strong general-purpose alternative tested; a code-tuned embedder is the open falsifier and runs through the same gate for free (content-hash cache; query-time weights).}
\label{tab:embed}
\end{table}

\bibliographystyle{unsrtnat}
\bibliography{refs}

\end{document}